\documentclass[aps,amsmath,amssymb,amsfonts,showpace,twocolumn,pra]{revtex4}

\usepackage{amscd}
\usepackage{amsthm}
\usepackage{graphicx}
\usepackage{mathdots}
\usepackage{rotating}
\usepackage{pgfplots}

\usepackage[
  colorlinks,
  linkcolor = blue,
  citecolor = blue,
  urlcolor = blue]{hyperref}

\usepackage{amsmath,amssymb,amsthm}

\usepackage{tikz}
\usetikzlibrary{arrows}
\usetikzlibrary{calc,positioning}

\usepackage{thm-restate}

\usepackage{colonequals}

\usepackage[font=small]{caption}

\newtheoremstyle{noCaption}
{\topsep}
{\topsep}
{\itshape}
{}
{}
{}
{0pt}
{}%

\def\qed{\hfill \vrule height7pt width 7pt depth 0pt}

\begin{document}
\title{The local indistinguishability of multipartite product states }
\author{Yan-Ling Wang$^{1}$, Mao-Sheng Li$^{2}$,   Zhu-Jun Zheng$^{1}$ and Shao-Ming Fei$^{3, 4}$}

 \affiliation
 {
 {\footnotesize  {$^1$Department of Mathematics,
 South China University of Technology, Guangzhou
510640, China}} \\
{\footnotesize  {$^2$Department of Mathematical Sciences,
 Tsinghua University, Beijing
100084, China}} \\
{\footnotesize{
  $^3$School of Mathematical Sciences, Capital Normal University,
Beijing 100048, China}}\\
{\footnotesize{$^4$Max-Planck-Institute for Mathematics in the Sciences, 04103
Leipzig, Germany}}
}

\begin{abstract}
 In this paper, we mainly study the local indistinguishability  of multipartite product states. Firstly, we follow the method of Z.-C. Zhang \emph{et al}[Phys. Rev. A 93, 012314(2016)] to give another more concise set of $2n-1$ orthogonal product states in $\mathbb{C}^m\otimes\mathbb{C}^n(4\leq m\leq n)$ which  can not be distinguished by local operations and classical communication(LOCC).  Then we use the 3 dimension cubes to present some product states which give us an intuitive view how to construct locally indistinguishable product states in tripartite quantum system. At last,  we give an explicit construction of locally indistinguishable orthogonal product states for general multipartite system.
\end{abstract}

\maketitle

\section{Introduction}
In quantum information theory, the problem that distinguishing the quantum states using local operations and classical communication (LOCC) has been extensively studied in the past 20 years, and numerous important results have been reported. The  local distinguishability of a given set of states is an important problem connected with the LOCC. In spite of these considerable efforts, the local indistinguishability of orthogonal multipartite states is
still incompletely solved.

Since the problem are considered, the maximally entangled states and the product states are concerned by most of the researchers. References \cite{Ghosh01,Wal02,Fan04,Nat05,Fan07,Yu12,Cos13,Yu14,Yu15}  are an incomplete list of  the results about the local distinguishability of maximally entangled states. Meanwhile, there are lots of people consider the local distinguishability of  product states \cite{Ben99,Ben991,Hor03,Rin04,Che04,Nis06,Fen09,Dua10,Yan13,Zha14,Ma14,Zhang15,Wang15,Zhang16}.
The product states was first considered by  Bennett. et al. who presented nine LOCC indistinguishable product states in $\mathbb{C}^3\otimes\mathbb{C}^3$ \cite{Ben99}. Since then, the locally indistinguishability
of orthogonal product states in the bipartite system have attracted much attention in recent years and have many advances\cite{Che04,Fen09,Zha14,Ma14,Zhang15,Wang15,Zhang16}. But for the multipartite system, there are a few
paper consider the locally indistinguishability of orthogonal product states\cite{Nis06,Dua10,Yan13}.  And in \cite{Zha14,Zhang15} give a full base of LOCC indistinguishable product states in tripartite system. Therefore, the study of LOCC indistinguishability of multipartite orthogonal product states is
still meaningful and interesting.

In this paper, we focus on finding the locally indistinguishable multipartite orthogonal product states. We separate it into two cases: the even parties and the odd parties.
For the even case, we first construct the bipartite system, then extend it into the any even parties. For  the odd cases, we first solve the tripartite case by considering the 3 dimensional cubic representation of the states which give an intuitive view for us to image such  states. Then we  combine it with the
even cases, we obtain the results for any odd parties.

We present our results in the following framework: In sec II, we present $2(n_2+n_4+\ldots+n_{2k}-k)+1$ LOCC indistinguishable orthogonal product states in even party system
$\mathbb{C}^{n_1}\otimes\mathbb{C}^{n_2}\otimes\cdots\otimes\mathbb{C}^{n_{2k}}$. In sec III, we first construct $2(n_1+n_3)-3$ orthogonal product states in tripartite system
$\mathbb{C}^{n_1}\otimes\mathbb{C}^{n_2}\otimes\mathbb{C}^{n_3}$, which cannot be perfectly distinguished by LOCC. Then we give $2(n_1+n_3+\ldots+n_{2k+1}-k)+1$ orthogonal product states
which are LOCC indistinguishable in $\mathbb{C}^{n_1}\otimes\mathbb{C}^{n_2}\otimes\cdots\otimes\mathbb{C}^{n_{2k+1}}.$

\section{constructions of even partite case}

\noindent {\bf Lemma. } In $\mathbb{C}^m\otimes\mathbb{C}^n(3\leq m\leq n),$ there are $2n-1$ LOCC indistinguishable orthogonal product states\cite{Zhang16}.

\begin{figure}[h]
\normalsize

\includegraphics[width=0.4\textwidth,height=0.22\textwidth]{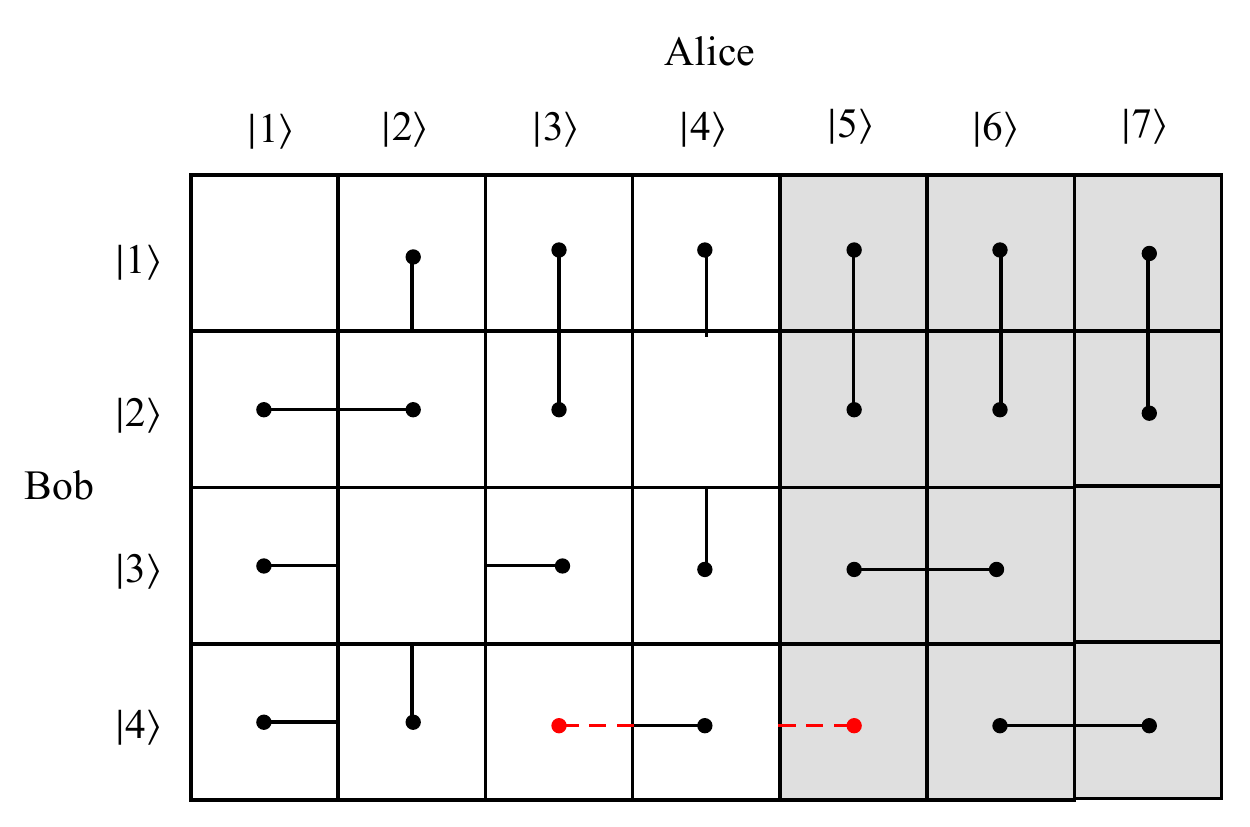}
  \caption{LOCC indistinguished product states in $\mathbb{C}^4\otimes\mathbb{C}^7.$}\label{bipartite_Example1}
\end{figure}

\begin{figure}[h]
\normalsize
\begin{sideways}
\includegraphics[width=0.265\textwidth,height=0.45\textwidth]{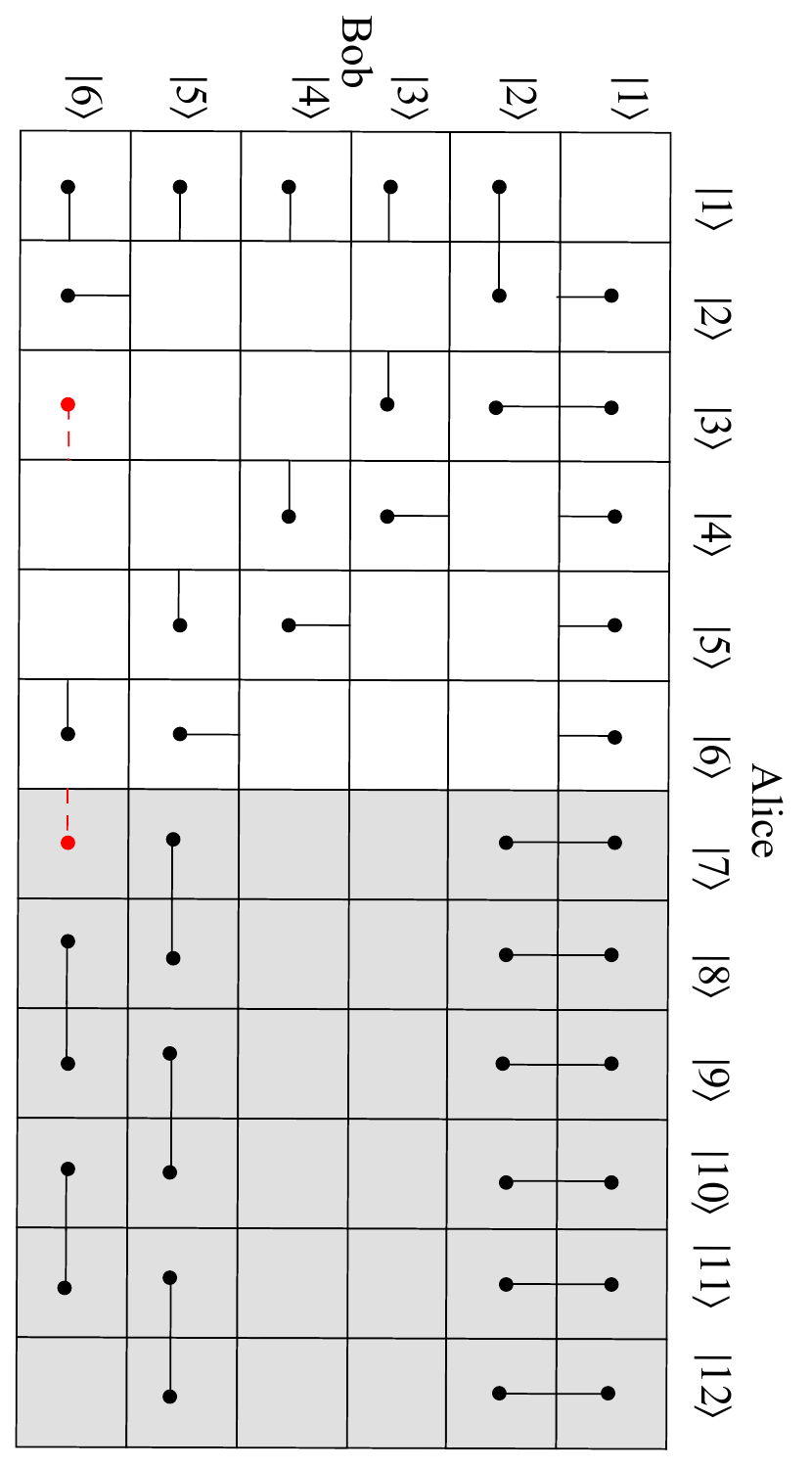}
\end{sideways}
\caption{LOCC indistinguished product states in $\mathbb{C}^6\otimes\mathbb{C}^{12}.$}\label{bipartite_Example2}
\end{figure}

Next, we give another set of product states which look much more concise than  the ones in lemma for $\mathbb{C}^m\otimes\mathbb{C}^n(4\leq m\leq n)$. Moreover, it is helpful for us to construct the tripartite  case. Firstly, we give the concrete construction as fllows:
$$
\begin{array}{l}
|\phi_1\rangle=(|1\rangle+|2\rangle+\cdots+|m\rangle)(|1\rangle+|2\rangle+\cdots+|n\rangle),\\
|\phi_i\rangle=|i\rangle|1-i\rangle, \ \ i=2,3,\ldots,m,\\
|\phi_{m+1}\rangle=|1-n\rangle|2\rangle,\\
|\phi_{m+j-1}\rangle=|1-(j-1)\rangle|j\rangle, \ \ j=3,4,\ldots,m,\\
|\phi_{m+l-1}\rangle=|1-2\rangle|l\rangle, \ \ l=m+1, m+2,\ldots, n,\\
|\phi_{m+n}\rangle=|m\rangle|3-(m+1)\rangle,\\
|\phi_{n+s}\rangle=|m-1\rangle|s-(s+1)\rangle,\\
|\phi_{n+t}\rangle=|m\rangle|t-(t+1)\rangle,\\
s=m+2k-1, t=m+2k, k=1,2,\ldots,\lfloor\frac{n-m}{2}\rfloor.
\end{array}
$$
Actually, we can show these states through the graph more clearly. We show two examples in Fig.\ref{bipartite_Example1} and Fig.\ref{bipartite_Example2}. We can see that the states in the left  white square are just as the states constructed in \cite{Zhang16,Yu15}. However, the states in the right grey rectangle are very simple and concise. The two red dashed lines constitute a product state.

Now, we show these quantum states are LOCC indistinguishable whose proof are very similar with the proof of the lemma.

\noindent {\bf Theorem 1.} In $\mathbb{C}^m\otimes\mathbb{C}^n(4\leq m\leq n),$ the above $2n-1$  orthogonal product states we constructed are LOCC indistinguishable.

\emph{Proof}. The proof are based on the following fact: if  some set of states are LOCC distinguishable, then some partite has to start with a nontrivial and nondisturbing measurement, ie. not all measurement $M^{\dagger}M$ are proportional to the identity and have the orthogonality relations preserved afterwards making further discrimination possible.

First, suppose Bob do the POVM measurement $M^{\dagger}_mM_m$, which is written as the following matrix
$$\left(
  \begin{array}{cccc}
    a_{11} &  a_{12} & \cdots&  a_{1m} \\
    a_{21} &  a_{22} & \cdots&  a_{2m} \\
    \vdots & \vdots & \ddots & \vdots \\
     a_{m1} &  a_{m2} & \cdots &  a_{mm} \\
  \end{array}
\right)
$$
in the basis $\{|1\rangle,|2\rangle,\ldots,|m\rangle\}.$

Since the postmeasurement states $(M_m\otimes I)|\phi_i\rangle$ are  orthogonal with each other.
Then we have $\langle\phi_j|M_m^{\dagger}M_m\otimes I|\phi_i\rangle=\langle j|\langle 1-j|M_m^{\dagger}M_m\otimes I|i\rangle|1-i\rangle=\langle j|M_m^{\dagger}M_m|i\rangle=a_{ji}=0, (i\neq j, i,j=2,3,\ldots,m).$ In the same way, consider the orthogonality $|\phi_i\rangle$ with $|\phi_{m+i}\rangle$ for $i=3,4,\ldots,m$ and $|\phi_{m+1}\rangle$ with $|\phi_2\rangle$, we have the $a_{1j}=0, j=2,3,\ldots,m$.

Consider the states $|\phi_1\rangle$ and $|\phi_{m+j}\rangle, j=1,2,\ldots,m$, we get $a_{11}=a_{22}=\cdots=a_{mm}.$ Thus $M_m^{\dagger}M_m$ are proportional to the identity.

When Alice start with the measurements, the matrix is denoted by $$M_n^{\dagger}M_n=\left(
  \begin{array}{cccc}
    b_{11} &  b_{12} & \cdots&  b_{1n} \\
    b_{21} &  b_{22} & \cdots&  b_{2n} \\
    \vdots & \vdots & \ddots & \vdots \\
     b_{n1} &  b_{n2} & \cdots &  b_{nn} \\
  \end{array}
\right)
$$
in the basis $\{|1\rangle, |2\rangle, \ldots, |n\rangle\}_B$

Similarly we have the postmeasurement states $(I\otimes M_n)|\phi_i\rangle$ are mutually orthogonal.
Consider the states $|\phi_{m+i}\rangle$ and $|\phi_{m+j}\rangle \ (i,j=1,2,\ldots,m+n-1)$, we can get $b_{ij}=0, i\neq j=2,3,\ldots,n.$
And the orthogonality $|\phi_i\rangle(i=2,3,\ldots,m)$ with $|\phi_{m+j-1}\rangle(j=2,3,\ldots,n)$, we have $b_{1j}=b_{ji}=0, j=2,3,\ldots,n$.
Finally, we consider the states $|\phi_1\rangle$ and $|\phi_{i}\rangle(i=2,3,\ldots,m,m+n,m+n+1,\ldots,2n-1)$, we get $b_{11}=b_{22}=\cdots=b_{nn}.$ Then we have the measurements $M^{\dagger}_nM_n$ are proportional to the identity.

Thus, these $2n-1$ product states are LOCC indistinguishable.\qed\\

To our surprise, the construction of small set of LOCC indistinguishable product states in even partite cases are more easy than the odd ones. Hence we first present the even cases.
\bigskip

\noindent {\bf Theorem 2. } In $\mathbb{C}^{n_1}\otimes\mathbb{C}^{n_2}\otimes\cdots\otimes\mathbb{C}^{n_{2k-1}}\otimes\mathbb{C}^{n_{2k}}(3\leq n_1\leq n_2\leq\cdots\leq n_{2k-1}\leq n_{2k}, k\geq 2) $ there exist $2(n_2+n_4+\cdots+n_{2k}-k)+1$ orthogonal product states which are LOCC indistinguishable.
The  construction of  the $2(n_2+n_4+\cdots+n_{2k}-k)+1$ product states are as follows:

Firstly,  we denote the product states   from the lemma  by $|\phi_{i_s}\rangle_s$ for the bipartite system $\mathbb{C}^{n_{2s-1}}\otimes\mathbb{C}^{n_{2s}},$ $ i_s=1,2,\ldots,2n_{2s}-2, s=1,2,\ldots,k.$  Then the multipartite product states are given by $k$ sets
$\{|\phi_{i_1}\rangle_1|11\rangle_2\cdots|11\rangle_k\},$ $\{|11\rangle_1|\phi_{i_2}\rangle_2|11\rangle_3\cdots|11\rangle_k\},$ $\cdots,$ $\{|11\rangle_1\cdots|11\rangle_{k-1} |\phi_{i_k}\rangle_k\}$
 where $(i_s=1,2\ldots2n_{2s}-2)$ with a stopper $$|\phi\rangle=\displaystyle\sum_{i_1=1}^{n_1}|i_1\rangle\displaystyle\sum_{j_1=1}^{n_2}|j_1\rangle\displaystyle
\cdots\displaystyle\sum_{i_k=1}^{n_{2k}}|i_k\rangle\displaystyle\sum_{j_k=1}^{n_{2k}}|j_k\rangle.$$
Next, we show these states are LOCC indistinguishable.

\emph{ Proof}: We call $\mathbb{C}^{n_{2s}-1}$ system the s-th Bob system.
$\mathbb{C}^{n_{2s}}$ system the s-th Alice system. When we consider the s-th Alice or Bob do the first measurement which must preserve the orthogonality of all the quantum states we presented. In particularly, it preserves the orthogonality of the following $2n_{2s}-1$ quantum states.

We use another notation to denote the states $|\psi_{i_s}\rangle=|11\rangle_1\cdots|11\rangle_{s-1}|\phi_{i_s}\rangle_s|11\rangle_{s+1}\cdots |11\rangle_{k}$ $(i_s=1,2,\ldots,2n_{2s}-2)$ and $|\psi_{2n_{2s}-1}\rangle=|\phi\rangle$, then the state $I\otimes\cdots\otimes I\otimes M_s^A\otimes I\cdots\otimes I|\psi_i\rangle$ orthogonal with the state $I\otimes\cdots\otimes I\otimes M_s^A\otimes I\cdots\otimes I|\psi_j\rangle$. After a  simple calculation, we obtain  $$_s\langle\phi_j| I\otimes M_s^{A^\dagger}M_s^{A}|\phi_i\rangle_s=\delta_{ij}.$$

By the proof of lemma,  any measurement which preserve the orthogonality of $|\phi_i\rangle_s$ and the corresponding stopper state must be trivial.  Then by the above formula,  we deduce that $M_s^{A^\dagger}M_s^{A}\propto I_s.$ Hence, in order to preserve the orthogonality of the above quantum states the only measurements of s-th Alice can do is the trivial measurement.

Similarly, if we consider  the s-th Bob system, we obtain $$_s\langle\phi_j | M_s^{B^\dagger}M_s^{B} \otimes I|\phi_i\rangle_s=\delta_{ij}.$$ Hence the measurement of the s-th Bob's system is also  the trivial measurement.
 \qed

\section{constructions of odd partite case}

The odd systems  cases are  different from the even ones. Firstly, we consider the simplest case: tripartite system. Then we combine the tripartite case with the even cases to tackle with the general odd cases.
Now, we consider a concrete example of odd partite.\\

\begin{figure}[h]
\normalsize
\includegraphics[width=0.5\textwidth,height=0.45\textwidth]{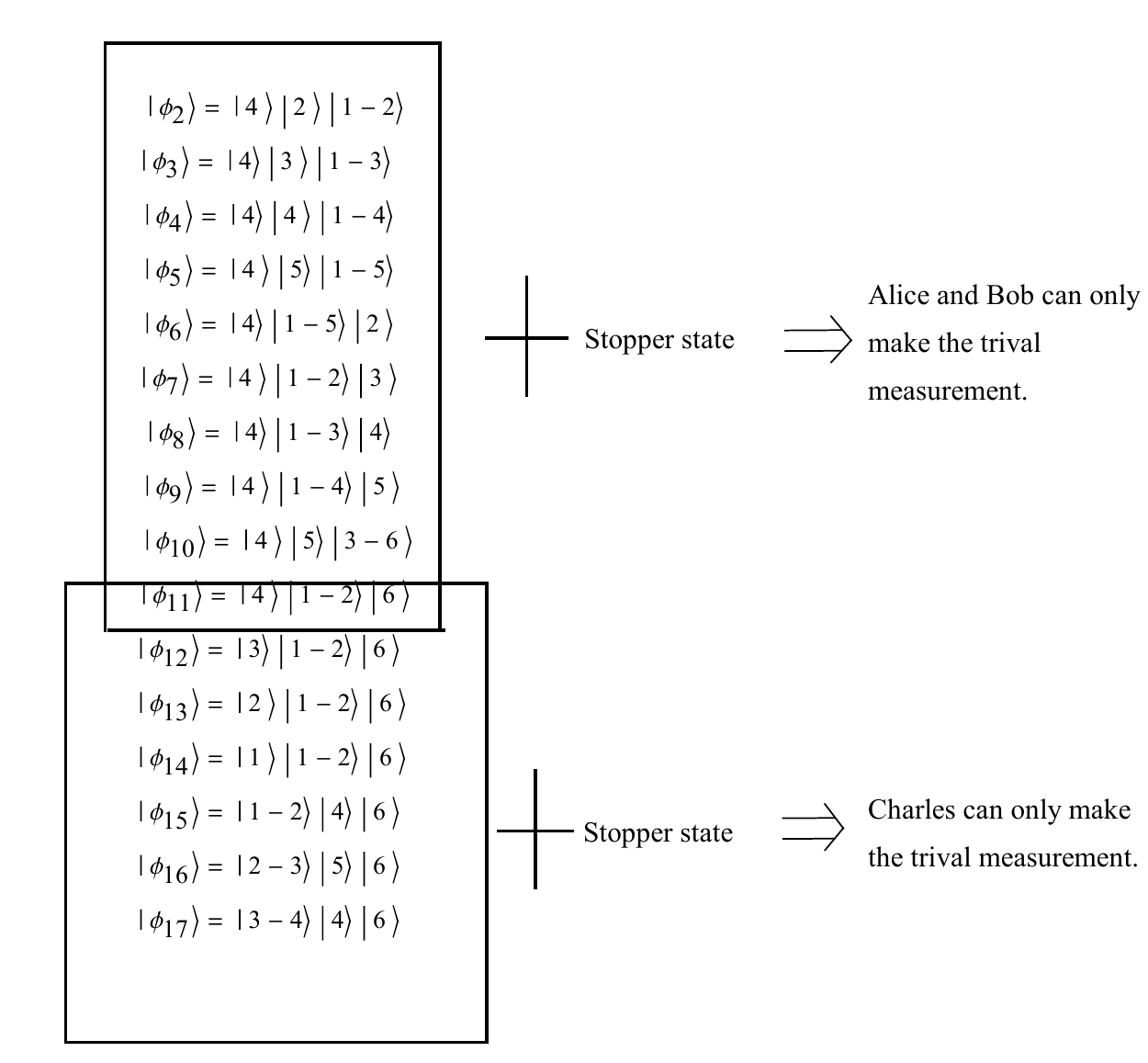}
  \caption{A sketch proof of the LOCC indistinguishability of states of example.}\label{simpleproof}

\end{figure}

{\noindent \bf Example. } In $\mathbb{C}^4\otimes\mathbb{C}^5\otimes\mathbb{C}^6$, there are 17 orthogonal product states that are LOCC indistinguishable.
Let
$$|\phi_1\rangle=|1+2+3+4\rangle|1+2+3+4+5\rangle|1+2+3+4+5+6\rangle$$
which is called a stopper state. By adding the stopper state into the set presented in Fig. \ref{simpleproof}, we obtain a set with  orthogonal product states in $\mathbb{C}^4\otimes\mathbb{C}^5\otimes\mathbb{C}^6$. Moreover, Fig. \ref{simpleproof} also gives us a sketch proof of the LOCC indistinguishability of this set.

\begin{figure}[h]
\normalsize

\includegraphics[width=0.44\textwidth,height=0.38\textwidth]{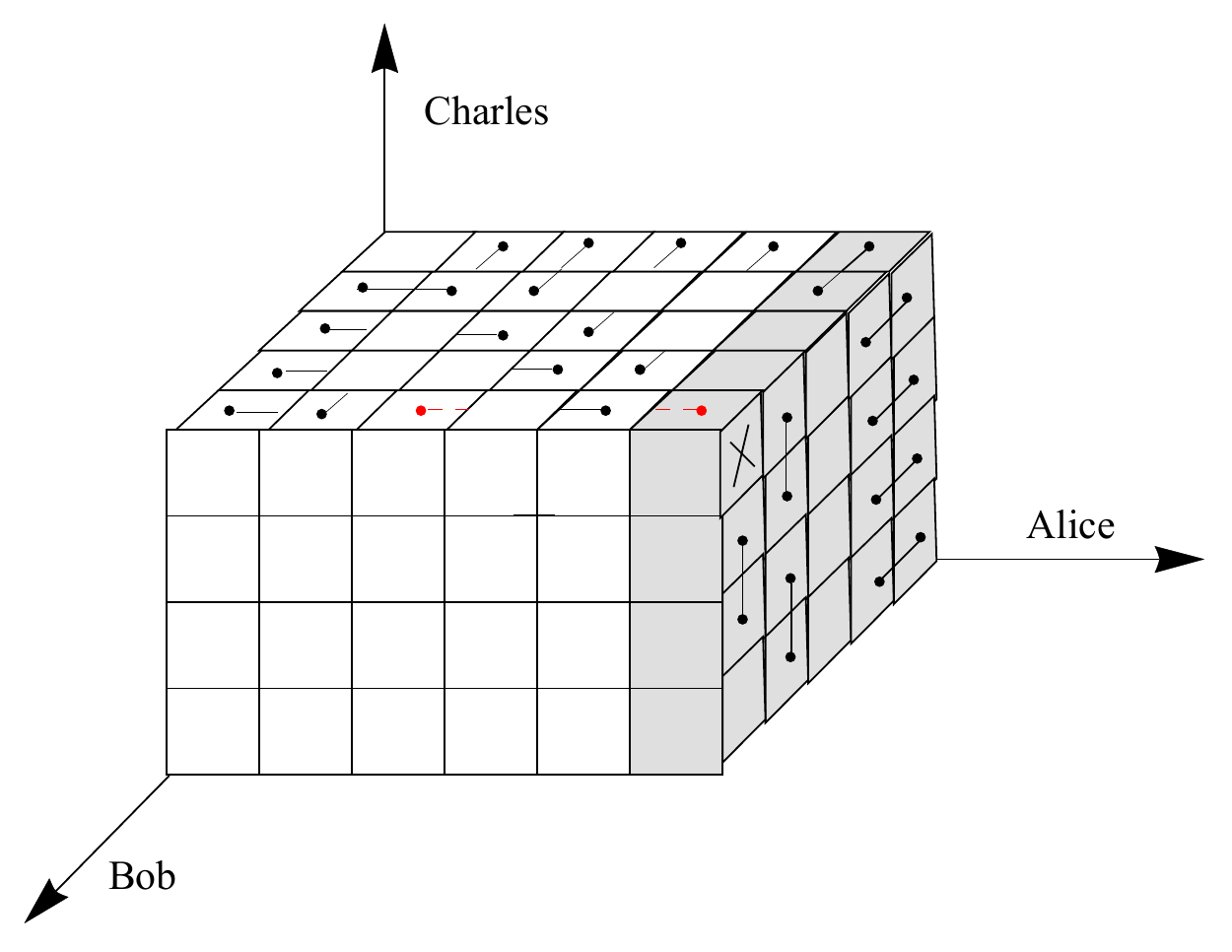}
 \caption{The representation of  states by  3 dimension cubes.}\label{cube}

\end{figure}

\begin{figure}[h]
\normalsize
\includegraphics[width=0.25\textwidth,height=0.20\textwidth]{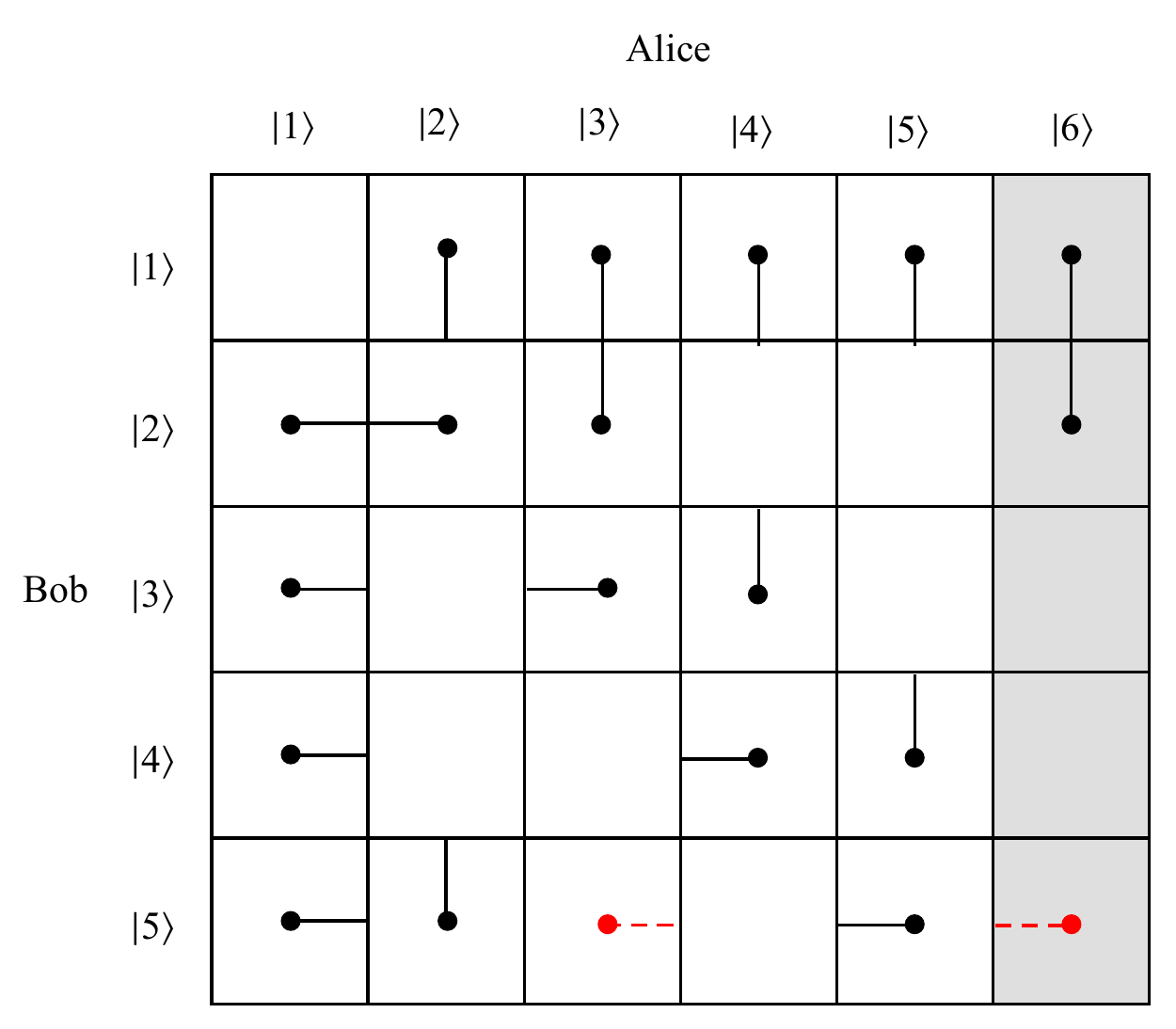}
\includegraphics[width=0.22\textwidth,height=0.17\textwidth]{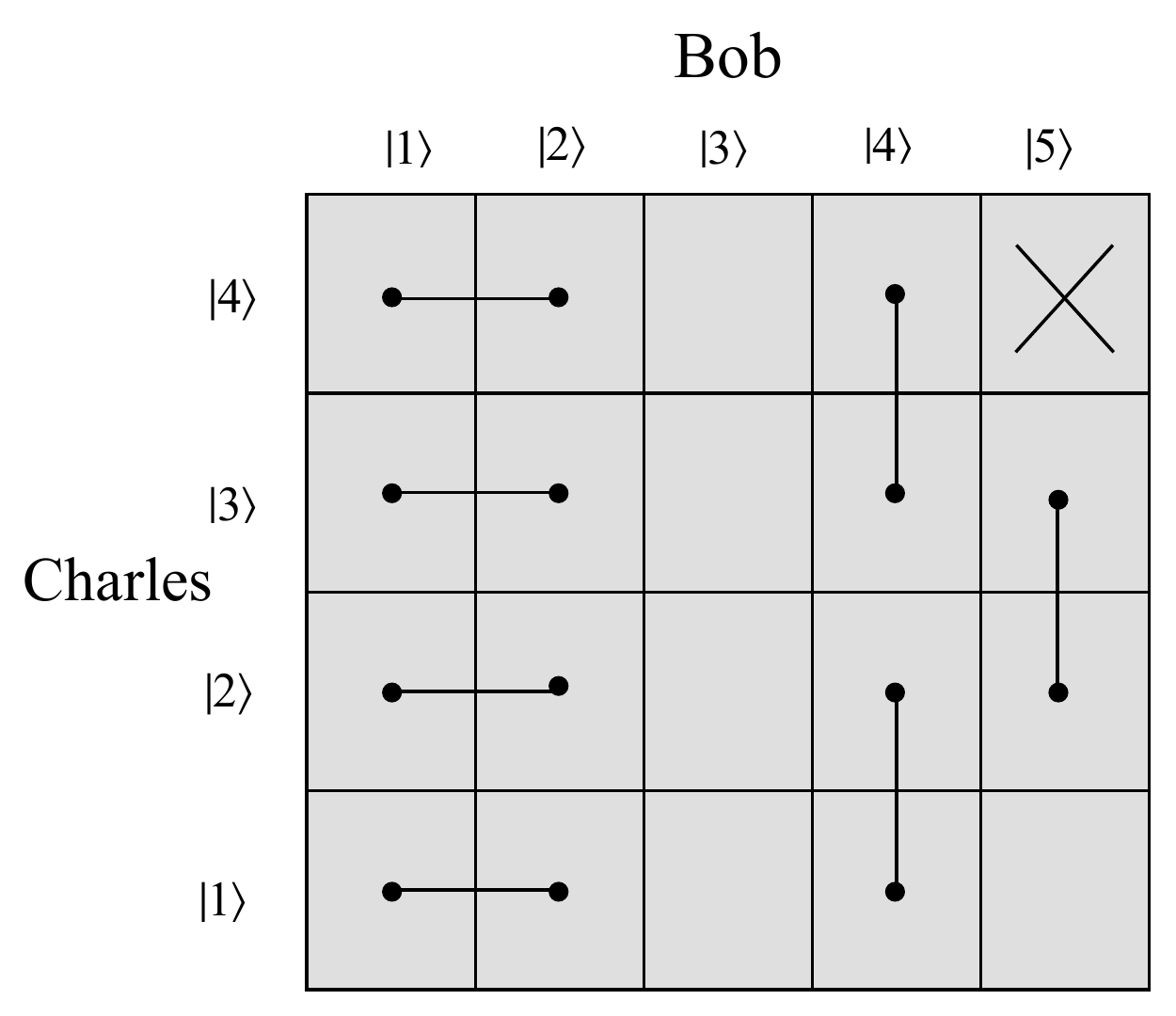}

 \caption{The top view of \textbf{Top surface part} and the left view of \textbf{Right surface part.}}\label{view}
\end{figure}

Similar with the bipartite case,  it might be intuitive to image a product state in tripartite quantum system  as some cube in three dimension space which is presented in Fig. \ref{cube}. Moreover, we give a projection view of this cubic representation in Fig. \ref{view}.

For the general tripartite system $\mathbb{C}^{n_1}\otimes\mathbb{C}^{n_2}\otimes\mathbb{C}^{n_3}$ $(4\leq n_1\leq n_2\leq n_3)$,  the orthogonal product states we want to  construct can be described by two parts: \textbf{Top surface part}  and \textbf{Right surface part} which is denoted by  $\mathcal{T},\mathcal{R}$ respectively.

 \textbf{Top surface part:}We denote the  $2n_3-1$  states except the stopper state in $\mathbb{C}^{n_2}\otimes\mathbb{C}^{n_3}$  constructed from theorem 1 by $\{|\psi_i\rangle\}_{i=1}^{2n_3-2}$. Then $\mathcal{T}=\{|\phi_i\rangle=|n_1\rangle\otimes|\psi_i\rangle \ \mid  \  i=1,2,...,2n_3-2\}.$ The number of elements in $\mathcal{T}$ is $2n_3-2$. Moreover, the top view of the states having been chosen is just the plan of dimension two case.

\begin{figure}[h]
\normalsize
\includegraphics[width=0.48\textwidth,height=0.28\textwidth]{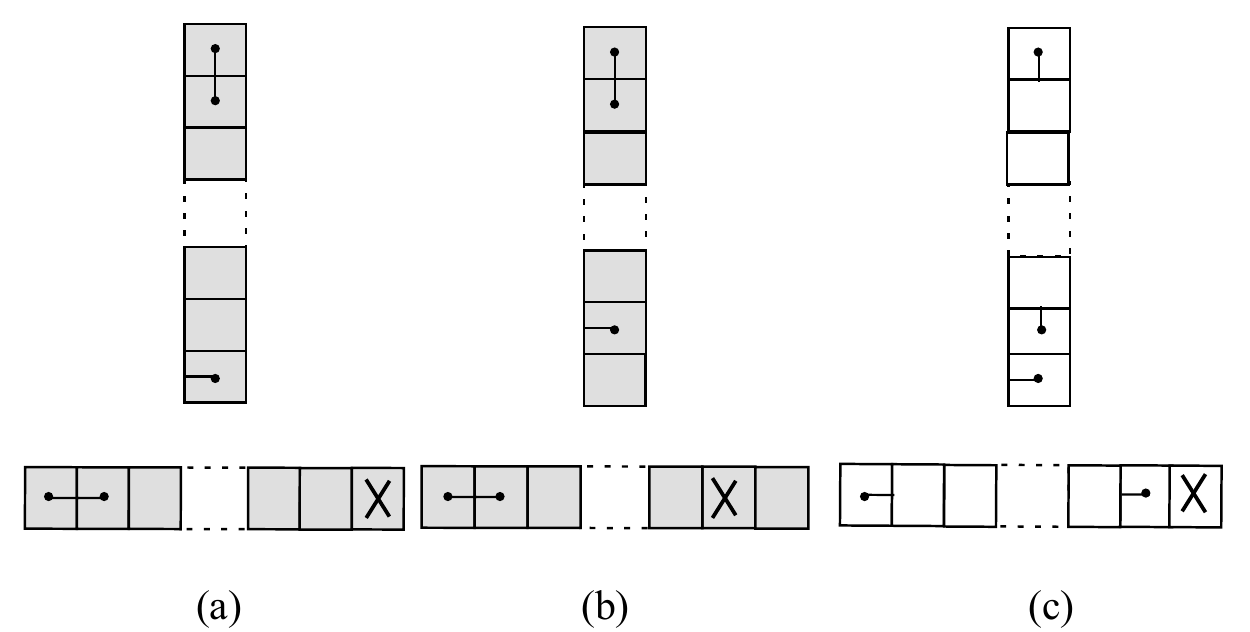}
 \caption{The cases of the right column of the top view and the corresponding left view of the \textbf{right surface part}.}\label{case}

\end{figure}

We notice that there are only three cases of the  last column of the top view: (a) $n_2<n_3$ and $n_3-n_2$ is odd, (a) $n_2<n_3$ and $n_3-n_2$ is even,(c) $n_2=n_3$. The corresponding views are showed in Fig. \ref{case}. Now the top row of the left view of the right surface also has three cases and are showed below the corresponding case in that figure. Now our aim is to add some states which might imply that Charles can only make a trivial measurements. In order to give an uniform description and make it more intuitive, we should have to permute the base of Bob (we denote the new base of Bob as $\{|i'\rangle\}_{i'=1}^{n_2}$). In case (a), there is no need to change. That is, $i'=i$ for all $i'$. In case (b), we should interchange $n_2-1$ and $n_2$. That is, $(n_2-1)'=n_2,$ $n_2'=n_2-1$, $i'=i$ for the other $i'$.  In case (c), we should interchange $2$ and $n_2-1$. That is, $2'=n_2-1,$ $(n_2-1)'=2$, $i'=i$ for the other $i'$. Now we can express the \textbf{Right surface part} more elegant under the new base of Bob. In fact,  under the new base of Bob, the left view of the right surface of the  states we are going to choose are just similar with the Fig.\ref{view}  or the grey part of Fig.\ref{bipartite_Example1} and \ref{bipartite_Example2}.

\textbf{Right surface part:} $\mathcal{R}$ is the union of some horizon states and some vertical states. The horizon part is $H$
$$H=\{|i\rangle|1'-2'\rangle|n_3\rangle \ \big| \ 1 \leq i\leq n_1\}.$$ The vertical part is $V$.
$$V=\{|i-(i+1)\rangle|(n_2-\delta_{(n_1-i)})'\rangle|n_3\rangle \ \big| 1 \  \leq i \leq n_1-1\}$$ where $\delta_i=\frac{1}{2}(1+(-1)^i)$, that is, $\delta_i=0$ if $i$ is odd and $\delta_i=1$ if $i$ is even.
Then $\mathcal{R}=H\cup V$.  Then number of elements in $\mathcal{R}$ is $2n_1-1$.

We notice that $\mathcal{R}\cap\mathcal{T}=\{|n_1\rangle|1'-2'\rangle|n_3\rangle \}$.  Then the set $\mathcal{S}$ of orthogonal product states we want to construct is adding the stopper state into $\mathcal{R}\cap\mathcal{T}$. Then the number of elements in $\mathcal{S}$ is $(2n_3-2)+(2n_1-1)-1+1$, that is,
$2(n_1+n_3)-3$. Here, the stopper state is $|\psi\rangle=|\psi_1\rangle_C|\psi_2\rangle_B|\psi_3\rangle_A$ where $|\psi_1\rangle_C=|1\rangle+|2\rangle+\ldots+|n_1\rangle, |\psi_2\rangle_B=|1\rangle+|2\rangle+\ldots+|n_2\rangle, |\psi_3\rangle_A=|1\rangle+|2\rangle+\ldots+|n_3\rangle.$ \\

\noindent {\bf Theorem 3. }In tripartite system $\mathbb{C}^{n_1}\otimes\mathbb{C}^{n_2}\otimes\mathbb{C}^{n_3}$ $(4\leq n_1\leq n_2\leq n_3)$, the set $\mathcal{S}$ we constructed above are a set with $2(n_1+n_3)-3$ orthogonal product states which is LOCC indistinguishable.

\noindent\emph{Sketch of proof:} If Alice or Bob applies a measurement that preserve the orthogonality of  \textbf{Top surface part} and the stopper state. After an easy calculation we may find that it is equivalent with that the corresponding measurement should preserve the orthogonality of the top view states and the stopper state of the reduced system without Charles. Then the measurement must be the trivial one.

If Charles applies a measurement that preserves the orthogonality of  \textbf{Right surface part} and the stopper state. Then his measurement must  also be the trivial one. Hence, none of the three people can apply a nontrivial measurement. \qed

Solving the tripartite system,  we can consider the general odd partite quantum system. \\

\noindent {\bf Theorem 4. } In $\mathbb{C}^{n_1}\otimes\mathbb{C}^{n_2}\otimes\cdots\otimes\mathbb{C}^{n_{2k}}\otimes\mathbb{C}^{n_{2k+1}}(4\leq n_1\leq n_2\leq\cdots\leq n_{2k+1})$, there are $2(n_1+n_3+\cdots+n_{2k+1}-k)+1$ product states that are LOCC indistinguishable.\\
\emph{Proof:} We give the proof by two steps: construction of states and proof of its LOCC indistinguishability.

Step 1:construction of states

Let $|\psi_{i_1}\rangle_1$  $i_1=1,2,\ldots,2(n_1+n_3)-4$ denote the product states except the stopper states we constructed in theorem 3 in $\mathbb{C}^{n_1}\otimes\mathbb{C}^{n_2}\otimes\mathbb{C}^{n_3}$. We call the corresponding system the 1-th Charles 1-th Bob and 1-th Alice system.

Let $|\psi_{i_s}\rangle_s$  $i_s=1,2,\ldots,2n_{2s+1}-2$ denote the product states except the stopper states we constructed in theorem 1 in $\mathbb{C}^{n_{2s}}\otimes\mathbb{C}^{n_{2s+1}}$. We call the corresponding system the s-th Bob and s-th Alice system for all integer $2\leq s\leq k$.

Then the set $\mathcal{S}$ of product states in multipartite quantum system we want to construct is the union of the following sets and a stopper state $|\psi\rangle$.
{\small
$$
\begin{array}{rcl}
  \mathcal{S}_1 & = & \{|\psi_{i_1}\rangle_1|11\rangle_2\cdots|11\rangle_k\  \big| \  i_1=1,2,\ldots,2(n_1+n_3)-4\}, \\
  \mathcal{S}_2 & = & \{|111\rangle_1|\psi_{i_2}\rangle_2|11\rangle_3\cdots|11\rangle_k  \big| \  i_2=1,2,\ldots,2(n_5)-2\}, \\
                & \vdots &                                                       \\
  \mathcal{S}_k & = & \{|111\rangle_1\cdots|11\rangle_{k-1}|\psi_{i_k}\rangle_k  \big| \  i_k=1,2,\ldots,2(n_{2k+1})-2\}.
\end{array}
$$
}
Here $|\psi\rangle=|\psi_1\rangle|\psi_2\rangle\ldots|\psi_{2k+1}\rangle$ where $|\psi_1\rangle=|1\rangle+|2\rangle+\cdots+|n_1\rangle,
|\psi_2\rangle=|1\rangle+|2\rangle+\cdots+|n_2\rangle,\cdots, |\psi_{2k+1}\rangle=|1\rangle+|2\rangle+\cdots+|n_{2k+1}\rangle.$ Then the number of elements in the set $\mathcal{S}$  is $2(n_1+n_3+\cdots+n_{2k+1}-k)+1$.

Now we need to check the orthogonality of the states in the set $\mathcal{S}=\mathcal{S}_1\cup \mathcal{S}_2\cup...\cup \mathcal{S}_k\cup\{|\psi\rangle\}$.
The orthogonality of the states in  $\mathcal{S}_s \cup \{|\psi\rangle\}$ is obvious by our construction for all $s=1,2,...,k$. So we only need to check that the states in $\mathcal{S}_i$ are orthogonal with the states in $\mathcal{S}_j$ whenever $i\neq j$. We observe that $|111\rangle_1$ is orthogonal with $|\psi_{i_1}\rangle_1$ for all  $i_1=1,2,\ldots,2(n_1+n_3)-4$ and $|11\rangle_s$ is orthogonal with $|\psi_{i_s}\rangle_s$ for all $i_s=1,2,...,2n_{2s+1}$ when $2\leq s\leq k.$ All these orthogonality give the orthogonality of the states between $\mathcal{S}_i$ and $\mathcal{S}_j$ for $i\neq j$.

Step 2: proof of its LOCC indistinguishability

If the 1-th Charles apply the first measurement, then his measurement should  preserve the orthogonal of the set of states $|\psi_{i_1}\rangle_1|11\cdots11\rangle$ and the stopper state. An easy calculation we obtain that it must be trivial one.

If the s-th Alice or Bob apply the first measurement, then his measurement should  preserve the orthogonal of the set of states $|11...11\rangle|\psi_{i_s}\rangle_s|11\cdots11\rangle$ and the stopper state. The same reason we obtain that it must be a trivial one. These complete the proof. \qed

\section{Conclusion and discussion}
In this paper, we first give a concise construction  of LOCC indistinguishable product states in $\mathbb{C}^m\otimes\mathbb{C}^n(4\leq m\leq n)$. Then we use the results of the lemma to present $2(n_2+n_4+\ldots+n_{2k}-k)+1$ LOCC indistinguishable orthogonal product states in even parties system
$\mathbb{C}^{n_1}\otimes\mathbb{C}^{n_2}\otimes\cdots\otimes\mathbb{C}^{n_{2k}}(3\leq n_1\leq n_2\leq\cdots\leq n_{2k})$. For the odd parties  system, we first consider the   simplest case: tripartite system. We use the 3 dimensional cubes  to give an intuitive view  to construct $2(n_1+n_3)-3$ orthogonal product states in tripartite system
$\mathbb{C}^{n_1}\otimes\mathbb{C}^{n_2}\otimes\mathbb{C}^{n_3}$, which cannot be perfectly distinguished by LOCC. At last, we give $2(n_1+n_3+\ldots+n_{2k+1}-k)+1$ orthogonal product states
are LOCC indistinguishable in $\mathbb{C}^{n_1}\otimes\mathbb{C}^{n_2}\otimes\cdots\otimes\mathbb{C}^{n_{2k+1}}(4\leq n_1\leq n_2\leq\cdots\leq n_{2k+1}).$

If we notice that   the simplest lower bound on the size of UPBs \cite{Div03,Alo01,Jia14} in $\mathbb{C}^{n_1}\otimes\mathbb{C}^{n_2}\otimes\cdots\otimes\mathbb{C}^{n_{k}}$ is $1+\sum_{i=1}^k(n_i-1).$ For the even parties cases with the same dimension in each system, that is, $n_1=n_2=...=n_{2k}=d$,  the bound $1+\sum_{i=1}^{2k}(n_i-1)$ is the same as the number $2(n_2+n_4+\ldots+n_{2k}-k)+1$ of LOCC indistinguishable orthogonal product states we presented in even parties system. However, the lower bound of the the size of UPBs  may not be reached, in which cases the local indistinguishability of  product states we presented is different from the local indistinguishability of UPBs. So it is interesting to compare with the two local indistinguishability.

\vspace{2.5ex}
\noindent{\bf Acknowledgments}\, \,
This work is supported by the NSFC 11571119, NSFC 11475178 and NSFC 11275131.


\begin{thebibliography}{}
\bibitem{Ghosh01}
S. Ghosh, G. Kar, A. Roy, A.Sen(De), and U. Sen,  Distinguishability of Bell States,
 Phys. Rev. Lett. \textbf{87}, 277902 (2001).

\bibitem{Wal02} J. Walgate and L. Hardy, Nonlocality, Asymmetry, and Distinguishing Bipartite States, Phys. Rev. Lett. {\bf 89}, 147901 (2002).



\bibitem{Fan04} H. Fan, Distinguishability and Indistinguishability by Local Operations and Classical Communication,  Phys. Rev. Lett. {\bf 92}, 177905 (2004).


\bibitem{Fan07} H. Fan, Distinguishing bipartite states by local operations and classical communication, Phys. Rev. A \textbf{75},014305 (2007).

\bibitem{Nat05}
M. Nathanson, Distinguishing bipartitite orthogonal states using LOCC: Best and worst cases, J. Math. Phys. {\bf 46}, 062103 (2005).


\bibitem{Yu12}
N. Yu, R. Duan, and M. Ying, Four Locally Indistinguishable Ququad-Ququad Orthogonal Maximally Entangled States, Phys. Rev. Lett. \textbf{109}, 020506
 (2012).

\bibitem{Cos13}
A. Cosentino, Positive partial transpose indistinguishable states via semidefinite programming,  Phys. Rev. A \textbf{87}, 012321 (2013).

 \bibitem{Yu14}
N. Yu, R. Duan, and M. Ying, Distinguishability of quantum states by positive operator-valued measures with positive partial transpose,
 IEEE Trans. Inf. Theory \textbf{60}, 2069
(2014).
\bibitem{Yu15} S. X. Yu and C. H. Oh, Detecting the local indistinguishability of maximally entangled states, arXiv:1502.01274v1.



\bibitem{Ben99}C. H. Bennett, D. P. DiVincenzo, C. A. Fuchs, T. Mor,E. Rains,
P. W. Shor, J. A. Smolin, and W. K. Wootters,  Quantum nonlocality without entanglement, Phys. Rev. A \textbf{59,}
1070 (1999).

\bibitem{Ben991}C. H. Bennett, D. P. DiVincenzo, T. Mor, P. W. Shor, J. A. Smolin, and B. M. Terhal,  Unextendible Product Bases and Bound Entanglement, Phys. Rev. Lett.  \textbf{82}, 5385 (1999).

\bibitem{Hor03}M. Horodecki, A. Sen(De), U. Sen, and K. Horodecki, Local Indistinguishability: More Nonlocality with Less Entanglement,
Phys. Rev. Lett. {\bf 90}, 047902 (2003).

\bibitem{Rin04} S. De Rinaldis, Distinguishability of complete and unextendible product bases,  Phys. Rev. A {\bf 70}, 022309(2004).

\bibitem{Che04} P.-X. Chen and C.-Z. Li, Distinguishing the elements of a full product basis set needs only projective measurements and classical communication,  Phys. Rev. A {\bf 70}, 022306 (2004).



\bibitem{Nis06} J. Niset and N. J. Cerf, Multipartite nonlocality without entanglement in many dimensions, Phys. Rev. A \textbf{74}, 052103 (2006).


\bibitem{Fen09}
Y. Feng and Y.-Y. Shi, Characterizing locally indistinguishable orthogonal product states, IEEE Trans. Inf. Theory \textbf{55}, 2799 (2009).



\bibitem{Dua10}
R. Y. Duan, Y. Xin, and M. S. Ying, Locally indistinguishable subspaces spanned by three-qubit unextendible product bases, Phys. Rev. A \textbf{81}, 032329 (2010).

\bibitem{Yan13}
Y.-H. Yang, F. Gao, G.-J. Tian, T.-Q. Cao, and Q.-Y. Wen, Local distinguishability of orthogonal quantum states in a $2\otimes 2\otimes 2$ system, Phys. Rev. A \textbf{88}, 024301 (2013).


\bibitem{Zha14} Z.-C. Zhang, F. Gao, G.-J. Tian, T.-Q. Cao  and Q.-Y. Wen, Nonlocality of orthogonal product basis quantum states, Phys. Rev. A {\bf 90}, 022313 (2014).

\bibitem{Ma14}T. Ma,   M.-J. Zhao, Y.-K. Wang and S.-M. Fei, Non-commutativity
and Local Indistinguishability of Quantum States. Sci. Rep. {\bf4}, 6336 (2014).

\bibitem{Zhang15}
  Z.-C. Zhang,  F. Gao,  S.J. Qin,  Y.-H. Yang and    Q.-Y.Wen, Nonlocality of orthogonal product states.
\newblock {\em Phys. Rev. A.} \textbf{91}, 012332 (2015).
\bibitem{Wang15}
 Y.-L.  Wang,   M.-S. Li,   Z.-J. Zheng and  S.-M. Fei, Nonlocality of orthogonal product-basis quantum states.
\newblock {\em Phys. Rev. A.} \textbf{92}, 032313 (2015).

\bibitem{Zhang16}
 Z.-C.  Zhang,   F. Gao,  Y. Cao,  S.J. Qin  and  Q.-Y. Wen: Local indistinguishability of orthogonal product states. \newblock {\em Phys. Rev. A.} \textbf{93}, 012314 (2016).


\bibitem{Div03}
D. P. DiVincenzo, T. Mor, P. W. Shor, J. A. Smolin, and B. M. Terhal, Unextendible product bases, uncompletable product bases and bound entanglement, Comm. Math. Phys. \textbf{238}, 379 (2003).

\bibitem{Alo01}N. Alon and L. Lov¨¢sz, unextendible product bases, J. Comb. Theor. Ser A \textbf{95}, 169 (2001).



\bibitem{Jia14} J. Chen, N. Johnston, The Minimum Size of Unextendible Product Bases in the Bipartite Case (and Some Multipartite Cases), Commun. Math. Phys.\textbf{ 333}, 351 (2015).




\end{thebibliography}
\end{document}